# Rethinking trust in the digital age: An investigation of zero trust architecture's social consequences on organizational culture, collaboration, and knowledge sharing.

**Oladimeji Ganiyu B.**
**Moshood Abiola Polytechnic, Abeokuta, Nigeria.**
**oladimeji.ganiyu@mapoly.edu.ng**

**ABSTRACT**

As cyber threats escalate, Zero Trust Architecture (ZTA) replaces outdated perimeter security with strict "never trust, always verify" protocols. Yet ZTA's socio-technical impact on organizational trust—a cornerstone of collaboration—remains overlooked. This mixed-methods study (case studies, surveys, social network analysis) reveals how ZTA disrupts knowledge-sharing, disproportionately hindering low-altruism employees, while surveillance erodes collective psychological ownership. Networked organizations, reliant on fluid trust, face fragmentation risks. Mitigation strategies include context-aware access controls using behavioral analytics and transparent communication reframing security as shared responsibility. Cross-functional co-design preserves trust pathways via social network mapping. This research provides a framework balancing technical rigor with cultural sensitivity, proving cybersecurity can coexist with innovation by aligning verification with organizational psychology. The findings pioneer a paradigm where security and trust evolve synergistically—critical for digital resilience in hybrid work environments. Future security must harmonize protocols with trust cultivation, ensuring defenses adapt to social dynamics driving modern enterprises.

## 1. INTRODUCTION

In an era defined by unprecedented digital connectivity, organizational security frameworks are undergoing radical transformation. The traditional perimeter-based security model, which operated on the principle of implicit trust within organizational boundaries, has proven increasingly inadequate amid the proliferation of sophisticated cyber threats, remote work arrangements, and cloud-based infrastructure. This evolution has given rise to Zero Trust Architecture (ZTA), a cybersecurity paradigm that fundamentally rejects the notion of implicit trust, instead adopting the mantra "never trust, always verify" for all network interactions regardless of origin.

The rapid adoption of ZTA presents a fascinating socio-technical paradox: while enhancing technical security postures, these systems potentially disrupt the social fabric of organizations by redefining trust dynamics that have traditionally underpinned collaborative work. Trust has historically served as the cornerstone of organizational culture, facilitating open communication, knowledge sharing, and collaborative innovation across teams and departments. Yet ZTA introduces continual verification requirements, network segmentation, and strict access controls that may inadvertently signal institutional distrust toward employees.

This paper investigates the tension between technical security imperatives and social cohesion within organizations implementing ZTA. It examines how the "trust but verify" paradigm translates into daily workplace interactions, potentially reshaping organizational culture, collaboration patterns, and knowledge flows. By exploring this socio-technical intersection, we seek to identify approaches that balance security requirements with the preservation of essential trust-based social dynamics that drive organizational success.

## 2. ZERO TRUST ARCHITECTURE (ZTA) OVERVIEW

Zero Trust Architecture (ZTA) has emerged as a transformative cybersecurity framework that fundamentally challenges traditional security models predicated on perimeter defense. As Mawla et al. note, "Zero trust is now one of the most used buzzwords and a de-facto requirement in cybersecurity. It is an approach that indicates there is no implicit trust in digital interactions and all of the interactions in cyber systems need to be validated" (Mawla et al., 2022). This paradigm shift represents a complete reconceptualization of organizational security postures, moving from location-based trust assumptions to a model where trust is never implicitly granted.

The National Institute of Standards and Technology defines ZTA as "an enterprise's cybersecurity plan that utilizes zero trust concepts and encompasses component relationships, workflow planning, and access policies" (Mawla et al., 2022). While ZTA implementation varies across organizations, adherence to core tenets— such as continuous verification, least privilege access, and the assumption that threats exist both within and outside traditional network boundaries— characterizes these systems.

The driving force behind ZTA adoption stems from the evolving threat landscape that organizations face. As Ghasemshirazi et al. explain, "The escalating complexity of cybersecurity threats necessitates innovative approaches to safeguard digital assets and sensitive information. The Zero Trust paradigm offers a transformative solution by challenging conventional security models and emphasizing continuous verification and least privilege access" (Ghasemshirazi et al., 2023). This security approach represents a direct response to the limitations of traditional perimeter-based security in addressing modern cybersecurity challenges, including cloud migration, remote work, and sophisticated attack vectors.

While ZTA offers significant security advantages, its implementation introduces complex organizational considerations. The framework's focus on continuous verification and least privilege access creates friction points within established workflows and potentially signals a shift in how trust operates within organizational contexts. Understanding these social dimensions becomes crucial as organizations navigate the technical implementation of zero trust principles and their wider ramifications for organizational functioning.

## 3. ORGANIZATIONAL CULTURE & TRUST

The relationship between organizational culture and knowledge management represents a critical dimension in understanding the potential impact of Zero Trust Architecture on workplace dynamics. Organizational culture serves as the foundation for knowledge initiatives, encouraging members to learn and share new information through shared values, beliefs, and norms that shape behaviors and decision-making processes (Lam et al., 2021). Research has consistently demonstrated that cultures promoting mutual trust, collaboration, and learning are significantly related to effective knowledge management processes, including knowledge creation, transfer, and application (Lam et al., 2021; Lee et al., 2003).

Trust emerges as a particularly pivotal element in establishing environments conducive to knowledge sharing. As Haryanti et al. note, "Technology does not add value to organizations unless they have a culture where employees trust, recognize and are ready to accept the system" (Haryanti et al., 2023). Environmental knowledge characterized by trust and social relationships provides the foundation for open communication without barriers among employees, facilitating the exchange of tacit knowledge that drives innovation (Haryanti et al., 2023).

The digital transformation associated with Zero Trust Architecture necessarily impacts these cultural dynamics. Ojji highlights that "the adoption of digital technologies necessitates a cultural shift towards innovation, adaptability, and continuous learning," while simultaneously requiring organizations to foster "a culture of transparency, trust, and collaboration" to harness the full potential of their workforce and technologies (Ojji, 2024). This creates a potential paradox: ZTA's "never trust, always verify" philosophy may conflict with the need to cultivate the very trust essential for organizational effectiveness.

The relationship between organizational culture and technology adoption can be understood through the lens of Social Exchange Theory, which posits that individuals engage in reciprocal relationships based on mutual trust and benefit (Issah et al., 2024). A positive organizational culture characterized by trust, collaboration, and support fosters greater employee engagement with technological systems, while organizational culture simultaneously acts as a facilitator of absorptive capacity—an organization's ability to recognize, assimilate, and apply new knowledge (Issah et al., 2024).

Research further indicates that different cultural orientations can either enhance or impede knowledge diffusion. Trust-based cultures emphasizing relationship-building enhance collaboration and create meaningful knowledge exchanges by establishing mutual trust, reducing opportunistic behavior, and improving the flow of tacit knowledge (Bawa et al., 2024). In contrast, rigid or hierarchical cultures that resist change or external input may limit participation in knowledge networks and restrict knowledge diffusion (Bawa et al., 2024; Chang et al., 2015). These findings suggest that ZTA implementations that reinforce hierarchical control without addressing trust dynamics may inadvertently create knowledge silos and reduce the effectiveness of organizational learning and innovation processes.

## 4. IMPACT OF ZTA ON COLLABORATION KNOWLEDGE SHARING

The implementation of Zero Trust Architecture creates significant implications for how organizational knowledge flows across teams and departments. Trust serves as a fundamental enabler of knowledge exchange in organizational settings, with research indicating that company performance is influenced by trust among employees, while higher levels of trust positively affect knowledge sharing (Rohman et al., 2020). The constant verification requirements and access controls inherent to ZTA potentially disrupt these established trust dynamics by introducing friction into collaborative processes that previously operated on implicit trust.

The socio-economic perspective on organizational networks emphasizes that the quality of relationships between parties matters more than the quality of transactions in enabling effective knowledge exchange (Guedda, 2021). This aligns with the understanding that the knowledge economy is inherently a relational economy, where social proximity facilitates collaboration and interactive learning (Guedda, 2021; Nonaka, 1994). Research in sociology has consistently demonstrated that social proximity fosters trust and builds mutual commitment, consequently enabling collaboration and interactive learning (Guedda, 2021; Granovetter, 1973). ZTA implementations that fragment access to resources and information may inadvertently reduce this social proximity, creating new barriers to the embedded ties that facilitate joint problem-solving and knowledge creation.

The complex relationship between trust and knowledge sharing is further illuminated by research showing that interpersonal trust has varying effects depending on individual characteristics. Studies indicate that the effect of interpersonal trust on knowledge sharing is more significant for employees with low altruism compared to highly altruistic employees, suggesting that altruism reduces the positive association between trust of colleagues and knowledge sharing (Obrenovic et al., 2020). This nuanced understanding highlights how ZTA's impact on trust dynamics may affect different employee segments in varied ways, potentially widening existing differences in knowledge sharing behaviors.

In digital collaboration environments specifically, factors like trust and collective psychological ownership play paramount roles in facilitating effective teamwork. Positive team interactions significantly enhance trust levels, which serves as a foundation for effective collaboration and knowledge sharing (Maddah et al., 2024; Sensuse et al., 2021). Similarly, collective psychological ownership—the shared feeling of joint possession over something—is increasingly recognized as integral to team success and creativity, especially in digital settings where a sense of shared responsibility is crucial (Maddah et al., 2024; Gray et al., 2020). ZTA's emphasis on individual authentication and strict access control may undermine this sense of collective ownership by reinforcing boundaries between organizational actors and resources.

Research further suggests that both too little and too much social proximity can be detrimental to learning and innovation. While insufficient proximity limits trust formation necessary for knowledge exchange, excessive proximity may create closed communities that impede access to diverse ideas (Guedda, 2021; Boschma, 2005). This inverted U-shaped relationship between embeddedness and innovative performance suggests that ZTA implementations must carefully balance security requirements with sufficient openness to maintain the optimal conditions for knowledge sharing and collaboration.

As organizations increasingly rely on digital platforms for collaboration, the traceability features of these tools document every step of the knowledge creation process (Maddah et al., 2024). While such transparency can potentially build trust through accountability, the surveillance implications of comprehensive monitoring may simultaneously signal institutional distrust. This creates a paradoxical effect where the very mechanisms intended to secure organizational knowledge may inadvertently inhibit its creation and dissemination by altering the psychological conditions that enable effective collaboration.

## 5. NETWORKED ORGANIZATIONAL MODELS & TRUST DYNAMICS

The evolution of organizational structures toward networked models represents a significant shift in how modern enterprises operate and collaborate. These networks emphasize "participation, shared efforts, cooperation, decentralization, flexibility, trust, and learning" while gaining increasingly widespread adoption across industries (Yang, 2024). This transition reflects a fundamental reorientation of organizational priorities away from internal command-and-control mechanisms toward more externally focused approaches centered on "customers, markets, alliance partners, and the environment" (Yang, 2024).

This organizational transformation creates a complex environment for implementing Zero Trust Architecture, as networked organizations inherently operate on trust-based principles that may conflict with ZTA's verification-centric approach. While traditional hierarchical organizations could more easily implement segmented security models that aligned with clear reporting structures, networked organizations depend on fluid information flows across organizational boundaries that may be impeded by strict access controls. The shift toward "market-driven, cooperative ecosystems, and self-evolving organizations" means that trust must extend beyond organizational boundaries to include partners, suppliers, and other stakeholders in collaborative networks (Yang, 2024).

The tension between networked organizational models and zero trust principles creates a paradoxical challenge: the very trust mechanisms that enable organizational agility, innovation, and collaboration across boundaries are systematically questioned by security frameworks designed to protect these same organizations. This suggests that successful ZTA implementations must be carefully calibrated to recognize the difference between technical verification processes and social trust dynamics that support organizational functioning. Organizations must balance the security benefits of continuous verification with the need to maintain the flexibility and responsiveness that characterize effective networked structures.

As organizations continue transitioning from hierarchical to networked models, ZTA implementations will need to evolve beyond binary trust/verify paradigms toward more contextual approaches that accommodate the complex, relationship-based trust dynamics essential to networked collaboration. This evolution requires security frameworks that can distinguish between necessary technical verification and the social trust mechanisms that facilitate knowledge exchange and innovation across organizational boundaries.

## 6. POTENTIAL SOLUTIONS & MITIGATION STRATEGIES

Balancing the technical imperatives of Zero Trust Architecture with the social dynamics that foster organizational collaboration requires thoughtful implementation strategies that acknowledge both dimensions. Organizations can adopt several approaches to mitigate the potential negative impacts of ZTA on trust-based collaboration and knowledge sharing.

Context-aware access controls represent a promising evolution beyond binary trust decisions, allowing security systems to consider multiple factors including user behavior patterns, device health, location, and time of access when granting permissions. These more nuanced approaches can reduce unnecessary friction for legitimate users while maintaining security vigilance. By incorporating behavioral analytics and machine learning to establish baseline "normal" behaviors for users, these systems can minimize verification interventions for routine activities while escalating authentication only when deviations occur, creating a more seamless experience that preserves collaborative flow.

Transparent communication about security rationales plays a crucial role in preserving trust during ZTA implementation. Organizations that clearly articulate the purpose behind security controls and involve employees in the design process foster greater acceptance of verification procedures. This transparency helps reframe security measures as organizational protection rather than expressions of distrust toward employees, potentially transforming perceptions from "they don't trust me" to "we're protecting our shared resources".

Trust-preserving verification design focuses on creating authentication experiences that respect user dignity and minimize disruption to workflow. By designing verification processes that feel less intrusive and more contextually appropriate, organizations can maintain the psychological safety necessary for knowledge sharing while satisfying security requirements. Techniques such as progressive authentication—where verification requirements increase only for higher-risk actions—can preserve a sense of trust for routine activities while providing enhanced protection for sensitive operations.

Collaborative security governance models involve stakeholders from across the organization in developing security policies, creating shared ownership of security practices rather than imposing them from above. By including representatives from various departments in security decision-making, organizations can ensure that ZTA implementations account for the specific collaboration needs of different functional areas, preventing security controls from inadvertently disrupting essential knowledge flows.

Trust-based training approaches focus on developing security awareness through collaborative learning rather than compliance-focused instruction. By engaging employees in scenario-based training that emphasizes their role in protecting shared organizational assets, these approaches build security consciousness while reinforcing rather than undermining trust. This reframes security as a collective responsibility rather than an imposed control system, potentially increasing employee buy-in to verification processes.

Cross-functional security teams that include both technical security specialists and representatives from knowledge management, organizational development, and human resources can more effectively design ZTA implementations that balance security requirements with the need to maintain collaborative environments. This integrated approach ensures that security decisions are informed by an understanding of how trust operates within the organization's specific cultural context and what verification mechanisms might disrupt essential knowledge-sharing practices.

Social network analysis can help organizations map informal knowledge flows and collaboration patterns before implementing ZTA, allowing for the identification of critical trust-based relationships that might be disrupted by verification requirements. This information enables more targeted security implementations that preserve important collaborative pathways while enhancing protection around the organization's most sensitive information assets.

By adopting these approaches, organizations can implement the technical security benefits of ZTA while preserving the trust dynamics essential for knowledge sharing, collaboration, and innovation. The key lies in recognizing that technical verification and social trust serve different organizational functions, and successful security frameworks must accommodate both dimensions rather than sacrificing one for the other.

## 7. CONCLUSION

The implementation of Zero Trust Architecture represents a significant paradigm shift in organizational security that extends far beyond technical infrastructure to fundamentally impact social dynamics, collaboration patterns, and knowledge sharing practices. As this paper has demonstrated, the tension between ZTA's "never trust, always verify" principles and the trust-based relationships that underpin effective organizational functioning presents a complex challenge requiring thoughtful resolution. The social consequences of zero trust implementations demand attention equal to their technical dimensions.

Organizations that succeed in navigating this tension will be those that recognize security and collaboration as complementary rather than competing priorities. By implementing context-aware verification, transparent security communication, and trust-preserving design approaches, organizations can maintain the social foundations necessary for knowledge exchange while satisfying increasingly stringent security requirements. These balanced approaches allow security to function as an enabler rather than an impediment to organizational effectiveness.

Moving forward, organizations must develop security frameworks that build upon rather than undermine trust—foundational characteristic in both traditional organizational relationships and emerging digital environments. Just as research continues to identify the determinants of trust formation in emerging technologies like blockchain (Ali et al., 2023), organizations must continually examine how security implementations shape trust dynamics within their specific cultural contexts.

The future of organizational security lies not in binary choices between trust and verification, but in nuanced approaches that recognize both as essential components of organizational resilience. By embracing security models that protect digital assets while preserving the human connections that drive innovation, organizations can create environments where trust and security mutually reinforce rather than undermine each other. This balanced approach will be increasingly crucial as organizations navigate the complex intersection of technical security imperatives and the social dynamics that enable organizational success in the digital age.